\documentclass[twocolumn,preprintnumbers,amsmath,amssymb]{revtex4-1}
\usepackage{graphicx}
\usepackage{dcolumn}
\usepackage{amsmath,amsbsy,amssymb,scalefnt}
\usepackage{bm}
\usepackage{epstopdf}
\usepackage{amsfonts}
\usepackage{textcomp}
\usepackage{mathtools}
\usepackage{hyperref}
\hypersetup{backref=true,
 pdfnewwindow=true, colorlinks=true,
 linkcolor=blue, anchorcolor=blue,
 citecolor=blue, filecolor=blue,
 menucolor=blue, urlcolor=blue}

\begin{document}


\title{Topological phase transition in GeSnH$_2$ induced by biaxial tensile strain: A tight-binding study}

\author{Zahra Aslani$^1$}
\author{Esmaeil Taghizadeh Sisakht$^{1,2}$}
\author{Farhad Fazileh$^2$}
\email{fazileh@cc.iut.ac.ir}
\author{H. Ghorbanfekr-Kalashami$^2$}
\author{F. M. Peeters$^2$}
\affiliation{$^1$Department of Physics, Isfahan University of Technology, Isfahan 84156-83111, Iran\\
$^2$Department of Physics, University of Antwerp, Groenenborgerlaan 171, B-2020 Antwerpen, Belgium}

\date{\today}

\begin{abstract}
An effective tight-binding (TB) Hamiltonian for monolayer GeSnH$_2$ is proposed which has an inversion-asymmetric honeycomb structure. The low-energy band structure of our TB model agrees very well with previous {\it ab initio} calculations under biaxial tensile strain. We predict a phase transition upon 7.5\% biaxial tensile strain in agreement with DFT calculations. Upon 8.5\% strain the system exhibits a band gap of 134 meV, suitable for room temperature applications. The topological nature of the phase transition is confirmed by: 1)the calculation of the $\mathbb{Z}_2$ topological invariant, and 2)quantum transport calculations of disordered GeSnH$_2$ nanoribbons which allows us to determine the universality class of the conductance fluctuations.
\end{abstract}

\maketitle


\section{\label{introduction}Introduction}

Topological insulators (TIs) have attracted a lot of attention in condensed matter physics and from the  materials science community
during the past decade~\cite{qi2011topological,hasan2010colloquium,moore2010birth,kane-mele2005,kanez22015}.  TIs are fascinating states of quantum matter with insulating bulk and topologically protected edge or
surface states. In two-dimensional (2D) TIs, also known as quantum spin Hall (QSH) insulators~\cite{kane-mele2005}, the gapless edge states
are topologically protected by time reversal symmetry (TRS) and are spin-polarized conduction channels that are robust against non-magnetic
scattering. They are more robust against backscattering than three-dimensional (3D) TIs, making them better suited for coherent transport
related applications, low-power electronics,  and quantum computing applications~\cite{chuang2014prediction}. 
 
There are currently only a few 3D TI compounds that are experimentally realized such as Bi$_2$Se$_3$, Bi$_2$Te$_3$,
and Sb$_2$Te$_3$~\cite{zhang2009topological}; and only HgTe/CdTe~\cite{konig2007quantum} and InAs/GaSb~\cite{knez2011evidence} quantum wells have been realized as 2D TIs. Also, due to the
very small bulk band gaps (on the order of meV), these 2D exhibit TI only at ultra-low temperatures. Therefore, there is a great need to find new 2D TIs with large energy band gaps. Following the advancements in graphene and similar materials, intensive efforts have been devoted to explore 2D group-IV and V honeycomb systems, which can harbor 2D topological phases. 
  
In terms of crystal structure, TIs can be generally divided into inversion-symmetric TIs (ISTIs) and inversion-asymmetric TIs (IATIs).
Inversion asymmetry introduces many additional intriguing properties in TIs such as pyroelectricity, crystalline-surface-dependent topological
electronic states, natural topological p-n junctions, and topological magneto-electric effects~\cite{ma2015,zhangh2016ydrogenated,li2016robust}.  Therefore, 2D IATI materials that are stable at room temperature would be highly promising candidates for future spintronics and quantum computing applications. 

Also, it would be interesting if we could implement such features in group-IV honeycomb systems for the integration of devices that use other carbon-IV honeycomb elemens. This would avoid issues such as contact resistances and their integration within the traditional Si or Ge based devices. 

An efficient method for the synthesis of suitable 2D nanomaterials with honeycomb structure is chemical functionalization. Hydrogenation and halogenation of the above systems for the realization of topological phases have been extensively explored. 

Here we consider GeSnH$_2$, which is an inversion asymmetric hydrogenated bipartite honeycomb system. {\it ab initio} calculations have shown that monolayers (ML) of GeSn halide (GeSnX$_2$, X$=$F, Cl, Br, I) are large band gap 2D TIs with protected edge states forming QSH systems~\cite{ma2015}. On the other hand hydrogenated ML GeSn (GeSnH$_2$) is a normal band insulator which can be transformed into a large band gap topological insulator via appropriate  biaxial tensile strain~\cite{ma2015}.  

Here, we propose a tight-binding (TB) model for a better understanding of the electronic band structure of GeSnH$_2$ near the Fermi level. The band structure of our TB model is fitted to the {\it ab initio} results, where we consider the cases with spin-orbit coupling (SOC) and without SOC. Within the linear regime of strain the band gap of our TB model agrees very well with the DFT results~\cite{ma2015}. This TB model when including SOC predicts a band inversion at 7.5\% biaxial tensile strain in agreement with DFT calculations~\cite{ma2015}. We show the topological nature of the phase transition by the calculation of the $\mathbb{Z}_2$ topological invariant. Quantum transport of this system is calculated in order to examine the protection of the edge states against nonmagnetic scattering. It is shown that the conductance fluctuations of disordered nanoribbons for energies near the band gap belong to the universality class of the circular unitary ensemble ($\beta=2$), while for high energies and strong disorder the fluctuations follow the circular orthogonal ensemble ($\beta=1$).     

This paper is organaized as follows. In Sec.~\ref{structure}, we introduce the crystal structure and lattice constants of  monolayer GeSnH$_2$. Our proposed TB model is introduced in 
Sec.~\ref{Model}, and the effect of strain on the electronic properties of  GeSnH$_2$ is examined. In Sec.~\ref{Phase}, the topological phase transition under strain is examined by looking at the nano-ribbon band structure and by determining the $\mathbb{Z}_2$ topological invariant. 
In Sec.~\ref{Transport} electronic transport in disordered GeSnH$_2$ nanoribbons is examined and the protection of the chiral edge states against nonmagnetic scatterings is verified. Also, we discuss the universality class of the conductance fluctuations. Our results are summarized in Sec.~\ref{Conclusion}.

\section{\label{structure}lattice structure}

\begin{figure}[ht] 
\centering
\vspace{20pt}
\centering
\includegraphics[width=.48\textwidth]{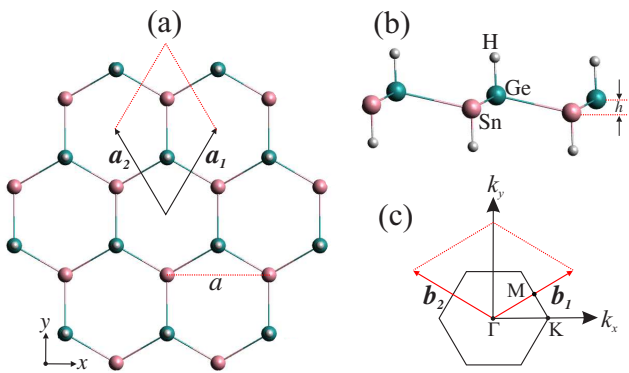}
\caption{Top (a) and side (b) views of the ML GeSnH$_2$ structure. ${\bm a_1}$ and ${\bm a_2}$ are the lattice vectors,
$a$ is the lattice constant and $h$ is the buckling height. (c) First Brillouin zone of the system with
three symmetry points $\Gamma$, K and M and the reciprocal
lattice vectors ${\bm b_1}$ and ${\bm b_2}$.}
\label{Lattice}
\end{figure}

The ML GeSnH$_2$ prefers a buckled honeycomb lattice, analogous
to its homogeneous counterparts germanene~\cite{germanene2011} and stanene~\cite{stanene2013}. 
Figs.~\ref{Lattice}(a) and \ref{Lattice}(b) show the atomic structure of ML GeSnH$_2$ and its geometrical parameters.
The 2D honeycomb lattice consists of two inequivalent sublattices
made of Ge and Sn atoms, which are named A and B sublattices, respectively.
Both Ge and Sn atoms exhibit $sp^3$ hybridization. One $sp^3$ orbital is passivated
by a hydrogen atom and the other three are bonded to three neighboring Ge or Sn atoms. Thus, the unit cell of GeSnH$_2$ consists of 
four atoms: one Ge, one Sn, and two Hydrogen atoms that are right above (below) the Ge (Sn) atoms.
The lattice translation vectors are
${\bm a}_{1,2}=a({\sqrt{3}}/{2}\hat{y}\pm{{1}/{2}}\hat{x})$ with the lattice constant $a=4.41$ \AA and the buckling height $(h)$ is $0.76$ \AA~\cite{ma2015,zhangh2016ydrogenated}.
Note that the $x$ and $y$ axes are taken to be along
the armchair and zigzag directions, respectively; and the $z$ axis is in the normal direction to the plane of the GeSnH$_2$ film. 

Fig.~\ref{Lattice}(c) shows the hexagonal Brillouin zone of ML GeSnH$_2$ with primitive reciprocal lattice vectors
${\bm b}_{1,2}={2\pi}/{a}({\sqrt{3}/3}\hat{y}\pm\hat{x})$ and three high symmetry points $\Gamma$, K, M.
The dynamical stability of ML GeSnH$_2$ was confirmed by the DFT calculated phonon spectrum~\cite{ma2015}.

\section{\label{Model}Low-energy effective TB Hamiltonian for monolayer G\lowercase{e}S\lowercase{n}H$_2$}

The electronic structure and topological properties of 2D honeycomb ML GeSnH$_2$ were studied in Ref.~\cite{ma2015} using first principle calculations based on DFT.
The DFT calculations including spin-orbit interaction predicted that a topological phase transition
is induced in ML GeSnH$_2$ through the application of biaxial in-plane tensile strain.
It was shown that the low-energy electronic structure of ML GeSnH$_2$ is
determined exclusively by using $s$, $p_{x}$ and $p_{y}$ atomic orbitals of Ge and Sn atoms~\cite{ma2015,zhangh2016ydrogenated}.
However, in order to examine the effect of random disorder on the electronic properties of this system and to confirm the topological nature of the phase transition, large system sizes are required. 
A limitation of the DFT calculations is that only small system sizes are managable. Therefore, we need a TB model for ML GeSnH$_2$ that is able to describe the low-energy electronic structure of large system sizes.

In the following we will derive a low-energy TB model including spin-orbit coupling (SOC) and we will show that the results
of the proposed TB model even in the presence of biaxial strain is in very good agreement with previous DFT calculations.
\begin{table*}
\caption{The first column shows the matrix elements for the nearest-neighbor hopping between $s$ and $p$ orbitals.
The hopping integrals as a function of direction-dependent quantities and SK parameters are listed in the second column.
The third column represents the hopping parameters with inclusion of applied strain.}
\label{table1}
\begin{ruledtabular}
\begin{tabular}{ccc}
\textrm{Hopping parameters}&
\textrm{Without strain}&
\textrm{With biaxial strain}\\
\colrule 
$t_{ss}$ & $V_{ss\sigma}$ & $t_{ss}^0[1-2\epsilon\cos^2\phi_0]$ \\ 
$t_{sp_x}$ & $lV_{sp\sigma}$  & $t_{sp_x}^0[1-2\epsilon\cos^2\phi_0+\eta\epsilon\tan\phi_0]$ \\
$\overline{t}_{sp_x}$ & $l\overline{V}_{sp\sigma}$ & $\overline{t}_{sp_x}^0[1-2\epsilon\cos^2\phi_0+\eta\epsilon\tan\phi_0]$ \\
$t_{sp_y}$ & $mV_{sp\sigma}$ & $t_{sp_y}^0[1-2\epsilon\cos^2\phi_0+\eta\epsilon\tan\phi_0]$  \\
$\overline{t}_{sp_y}$ & $m\overline{V}_{sp\sigma}$ & $\overline{t}_{sp_y}^0[1-2\epsilon\cos^2\phi_0+\eta\epsilon\tan\phi_0]$\\
$t_{p_xp_x}$&$l^2V_{pp\sigma}+(1-l^2)V_{pp\pi}$ & $t_{p_xp_x}^0[1-2\epsilon\cos^2\phi_0+2\eta\epsilon\tan\phi_0]-2\eta\epsilon\tan\phi_0V_{pp\pi}$ \\
$t_{p_yp_y}$&$m^2V_{pp\sigma}+(1-m^2)V_{pp\pi}$ & $t_{p_yp_y}^0[1-2\epsilon\cos^2\phi_0+2\eta\epsilon\tan\phi_0]-2\eta\epsilon\tan\phi_0V_{pp\pi}$ \\
$t_{p_xp_y}$&$lm(V_{pp\sigma}-V_{pp\pi})$ & $t_{p_xp_y}^0[1-2\epsilon\cos^2\phi_0+2\eta\epsilon\tan\phi_0]$\\ 
\end{tabular}
\end{ruledtabular}
\end{table*}

\subsection{\label{SOC}Tight-Binding model Hamiltonian without SOC}
To describe the low-energy spectrum and the electronic properties of ML GeSnH$_2$, we propose a TB model hamiltonian involving the three
outer-shell $s$, $p_x$ and $p_y$ atomic orbitals of Ge and Sn atoms. The effective TB Hamiltonian without SOC can be written in the second quantized
representation as
\begin{equation}
\begin{aligned}
H_{0}&=\sum_{i\alpha}E_{i\alpha} c_{i\alpha}^{\dag} c_{i\alpha}  \\
&+\sum _{\langle i,j\rangle ;\alpha ,\beta}t_{i\alpha ,j\beta}(c_{i\alpha}^{\dag} c_{j\beta} +h.c.).
\end{aligned}
\label{hamiltoni}
\end{equation}
Here, $\alpha$, $\beta\in (s,p_x,p_y)$ are the orbital indices and $\langle i,j\rangle$ denotes the nearest-neighboring $i$th
and $j$th atoms. $E_{i\alpha}$ is the on-site energy of $\alpha$th orbital of $i$th atom,
$c_{i\alpha}^{\dag} (c_{i\alpha})$ is the creation (annihilation) operator
of an electron in the $\alpha$th orbital of the $i$th atom, and $t_{i\alpha ,j\beta}$ is the nearest-neighbor hopping parameter
between $\alpha$th orbital of $i$th atom and $\beta$th orbital of $j$th atom. 
\begin{table}
\caption{Numerical values of the SK parameters obtained from a fitting to the $ab$ $initio$ results. The energy units are eV.}
\label{table2}
\begin{ruledtabular}
\begin{tabular}{ccccccccc}
\textrm{$V_{ss\sigma}$}&
\textrm{$V_{sp\sigma}$}&
\textrm{$\overline{V}_{sp\sigma}$}&
\textrm{$V_{pp\sigma}$}&
\textrm{$V_{pp\pi}$}&
\textrm{$E_{Ge,s}$}&
\textrm{$E_{Ge,p}$}&
\textrm{$E_{Sn,s}$}&
\textrm{$E_{Sn,p}$}\\
\colrule
-1.51&3.33 &2.35&3.69&-1.03&-5.47&5.23&-2.26&1.91
\end{tabular}
\end{ruledtabular}
\end{table}
The hopping parameters of Eq.~(\ref{hamiltoni}) are determined by the Slater-Koster (SK)~\cite{slater-k1954}
integrals as shown in the second column of table~\ref{table1}, where $l=\cos\theta\cos\phi _0$
and $m=\sin\theta\cos\phi _0$ are direction cosines of the angles of the vector connecting
two nearest-neighboring atoms with respect to $x$ and $y$
axes, respectively.

We calculated the hopping parameters and on-site energies of the above Hamiltonian using the method of minimization of the least square
difference between the DFT obtained band structure based on the Heyd-Scuseria-Ernzerhof (HSE) approximation~\cite{ma2015} and the band structure of our
TB model.
Our TB model Hamiltonian has nine fitting parameters; namely, four on-site orbital
energies $(E_{Ge,s}, E_{Ge,p}, E_{Sn,s}, E_{Sn,p})$
and five SK parameters related to the hopping energies $(V_{ss\sigma}, V_{sp\sigma}, \overline{V}_{sp\sigma}, V_{pp\sigma}, V_{pp\pi})$.
Note that $V_{sp\sigma} (\overline{V}_{sp\sigma})$ is the hopping integral between $s$ orbitals of atoms in sublattice A (B) and
$p$ orbitals of atoms in sublattice B (A).

\begin{figure}[ht] 
\centering
\vspace{20pt}
\includegraphics[width=0.47\textwidth]{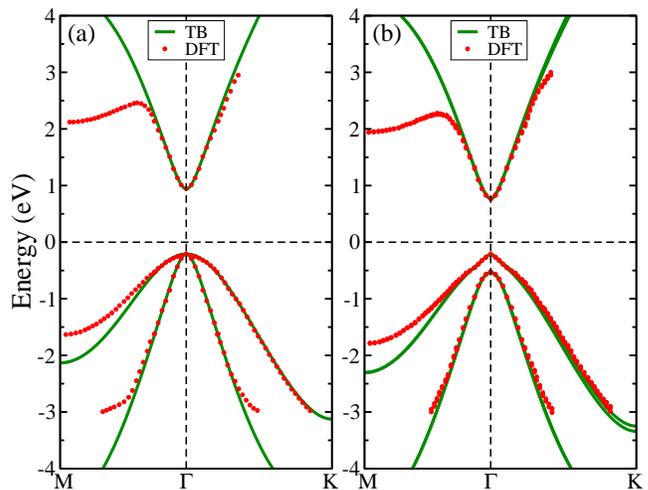}
\caption{The TB band structure of ML GeSnH$_2$ (a) without SOC and (b) with SOC.
Symbols represent the DFT-HSE data taken from Ref.~\cite{ma2015}.
}
\label{bands1}
\end{figure}

Table~\ref{table2} presents the obtained numerical values of the SK parameters. Using the optimized
parameters, we can reproduce the three low-energy bands near the Fermi level ($s$ and $p_{xy}$ bands). Fig.~\ref{bands1}(a) shows the TB low-energy bands of ML GeSnH$_2$ that
is in good agreement with the DFT results. The band structure has a direct band gap of 1.155~eV at the $\Gamma$ point.

\subsection{\label{SOC}Spin-orbit coupling in ML G\lowercase{e}S\lowercase{n}H$_2$}
In general, spin-orbit interaction can be written as~\cite{liu2011low}
\begin{equation}
H_{SOC}=\frac{\hbar}{4m_0^2c^2}(\bm{\nabla} V\times\bm{p})\cdot\bm{\sigma},
\label{HSOC}
\end{equation}
where, $\hbar$ is Planks constant, $m_0$ is the rest mass of an electron, $c$ is the velocity of light, $V$ is
potential energy, $\bm{p}$ is momentum, and $\bm{\sigma}$ is the vector of Pauli matrices. The major part
of SOC in systems that consist of heavy atoms comes from the orbital motion of electrons close to the atomic nuclei.
In such systems and within the central field approximation, the crystal potential $V(\bm{r})$ can be considered as an effective
spherical atomic potential $V_i(\bm{r})$ located at $i$th atom.
Therefore, by substituting $\bm{\nabla} V_i(\bm{r})=({{dV_i}/{dr}}){{\bm r}/{r}}$ and $\bm{s}=({\hbar}/{2})\bm{\sigma}$
terms into Eq.~(\ref{HSOC}) the SOC term takes the form~\cite{liu2011low},
\begin{equation}
H_{SOC}=\lambda (r)\bm{L}\cdot\bm{s}.
\label{HSOC1}
\end{equation}
The above equation can also be expressed in the form
\begin{equation}
H_{SOC}=\lambda (r)\bigg(\frac{L_+s_-+L_-s_+}{2}+L_zs_z\bigg),
\label{HSOC2}
\end{equation}
where, $\lambda (r) ={1}/{2m_o^2c^2}({1}/{r})({dV}/{dr})$ is the  effective atomic SOC constant whose
value depends on the specific atom.
$L_{\pm}$, $s_{\pm}$ are the operators for angular momentum and spin, respectively.
In the basis set of
$\lvert s_{A}, p_{xA}, p_{yA}, s_{B}, p_{xB}, p_{yB}\rangle\otimes\lvert\uparrow ,\downarrow\rangle$ ,
the matrix elements of the on-site SOC Hamiltonian for ML GeSnH$_2$ are given by 
\begin{equation}
\langle \alpha _i\vert H_{SOC}\vert\beta _i\rangle=\lambda _i\langle\frac{L_+s_-+L_-s_+}{2}+L_zs_z\rangle_{\alpha\beta}.
\end{equation}
Here $\alpha _i$ and $\beta _i$ are the atomic orbitals and $\lambda _i$ is the on-site SOC strength of the $i$th atom.
Note that since the two atoms in the unit cell of ML GeSnH$_2$ are different, we have two distinct SOC strengths 
$\lambda _A$ and $\lambda _B$ for atoms in sublattice A (Ge atoms) and sublattice B (Sn atoms), respectively.\\
The resulting SOC Hamiltonian matrix in the above basis is given by
\begin{equation}
H_{SOC}=\left(\begin{array}{cc}
H_{SOC}^{\uparrow\uparrow} & H_{SOC}^{\uparrow\downarrow} \\
H_{SOC}^{\downarrow\uparrow}  & H_{SOC}^{\downarrow\downarrow}\\
\end{array}\right),
\end{equation}
where, the elements are 6$\times$6 matrices
\begin{equation}
H_{SOC}^{\uparrow\uparrow}=\frac{1}{2}\left(\begin{array}{cccccc}
0 & 0 & 0 & 0 & 0 & 0 \\
0 & 0 & -i\lambda_{A} & 0 & 0 & 0\\
0 & i\lambda_{A} & 0 & 0 & 0 & 0\\
0 & 0 & 0 & 0 & 0 & 0 \\
0 & 0 & 0 & 0 & 0 & -i\lambda_{B}\\
0 & 0 & 0 & 0 & i\lambda_ {B} & 0 \\
\end{array}\right),
\end{equation}
\begin{equation}
\begin{split}
& H_{SOC}^{\downarrow\downarrow}=-H_{SOC}^{\uparrow\uparrow}~,\\
& H_{SOC}^{\uparrow\downarrow}=H_{SOC}^{\downarrow\uparrow}=0~.
\end{split}
\end{equation}

The value of the on-site effective SOC strength $\lambda$ is determined by fitting 
the TB energy bands to the DFT results. 
Here we fitted the energy bands obtained from our TB model to the one from the DFT+HSE approach~\cite{ma2015}.
The optimized numerical results using the hopping parameters from Table~\ref{table2} are $\lambda _{A}=0.296$~eV,
$\lambda _{B}=0.326$~eV for Ge and Sn atoms, respectively. 
The TB energy bands of ML GeSnH$_2$ in the
presence of spin-orbit interaction using the mentioned values of SOC strengths are in good agreement with 
the $ab$ $initio$ results as shown in Fig.~\ref{bands1}(b).

Notice from Fig.~\ref{bands1}(b) the spin splitting in the doubly degenerated bands. The splitting is more clear for the upper valence band close to the K symmetry point which is 92 meV. For the conduction band the spin-splitting is 241 meV at the K symmetry point. All the energy states of a system with inversion symmetry will be spin-degenerate while the TRS is held. 
The degeneracy can be lifted by breaking the inversion symmetry in such a system.
The band gap of the system is reduced to 0.977 eV, demonstrating that
ML GeSnH$_2$ is still a normal semiconductor. 

\subsection{\label{S&E}The effect of strain}
The electronic properties of a system can be affected significantly by applying strain~\cite{bir1974,sun2010}. This is due to the fact that
strain changes both the bond lengths and the bond angles. This in turn changes the SK parameters and hopping integrals that further affect the electronic band structure. 

Based on knowledge from our previous works, we expect a topological phase transition upon the application of biaxial tensile strain~\cite{ma2015}. Further, since the electronic band structure of this system under strain is available from $ab$ $initio$ calculations, the comparison of our TB band structure with DFT results will be another verification for the correctness of our TB model Hamiltonian.

Here, we investigate the effect of biaxial tensile strain on ML GeSnH$_2$. In our previous work we studied the effect of biaxial tensile strain on a related system where the low-energy electronic properties were dominated by $s$, $p_x$ and $p_y$ atomic orbitals~\cite{rezaei2017GeCH3}. We follow the approach of Ref.~\cite{rezaei2017GeCH3} and obtain the effect of strain on the hopping parameters which we list in table~\ref{table1}. Note that $t_{i\alpha,j\beta}^{0}$ indicates the unstrained hopping parameters.

In such systems, it is expected that the buckling angle changes linearly with strain as $\phi =\phi_0 -\eta\epsilon$~\cite{rezaei2017GeCH3}. $\epsilon$ is the strength of the applied biaxial strain and $\phi_0$ is the initial buckling angle.
However, our calculations show that the changes in the buckling angle in this system produces negligible effect on the hopping parameters and therefore, we will use the initial buckling angle $\phi_0$ in the rest of our calculations.

\begin{figure}[ht]
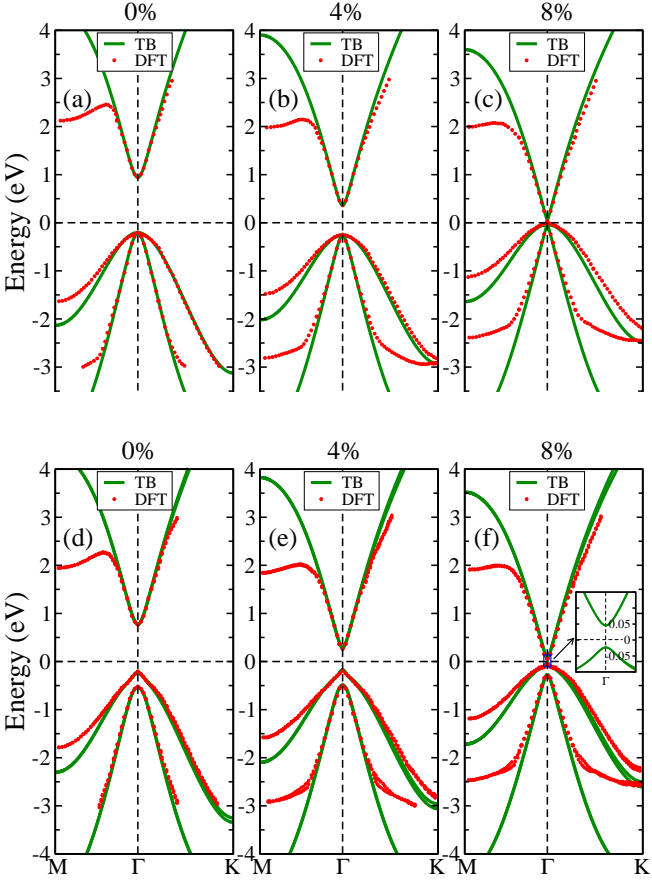
 
\centering
\vspace{20pt}
\includegraphics[width=0.48\textwidth]{bands2_ab.eps}
\includegraphics[width=0.48\textwidth]{bands2_SOC_cd.eps}
\caption{The TB band structure of ML GeSnH$_2$ without (upper panels) and with (lower panels) SOC in the presence of (a) and (d) 0\%,
(b) and (e) 4\%, and (c) and (f) 8\% biaxial tensile strain.
Red dots indicate the DFT-HSE data taken from Ref.~\cite{ma2015}.}
\label{bands2}
\end{figure}

The Hamiltonian for the strained system can be obtained by substituting the new hopping parameters
(last column of table~\ref{table1}) in the original Hamiltonian Eq.~(\ref{hamiltoni}). 
The calculated TB energy spectrum of ML GeSnH$_2$ are shown in Figs.~\ref{bands2}(a-c) without SOC and Figs.~\ref{bands2}(d-f) with SOC
in the presence of 0\%, 4\%, and 8\% biaxial tensile strains. As shown, these results are in good agreement with the DFT calculations~\cite{ma2015}.

\begin{figure}[ht]
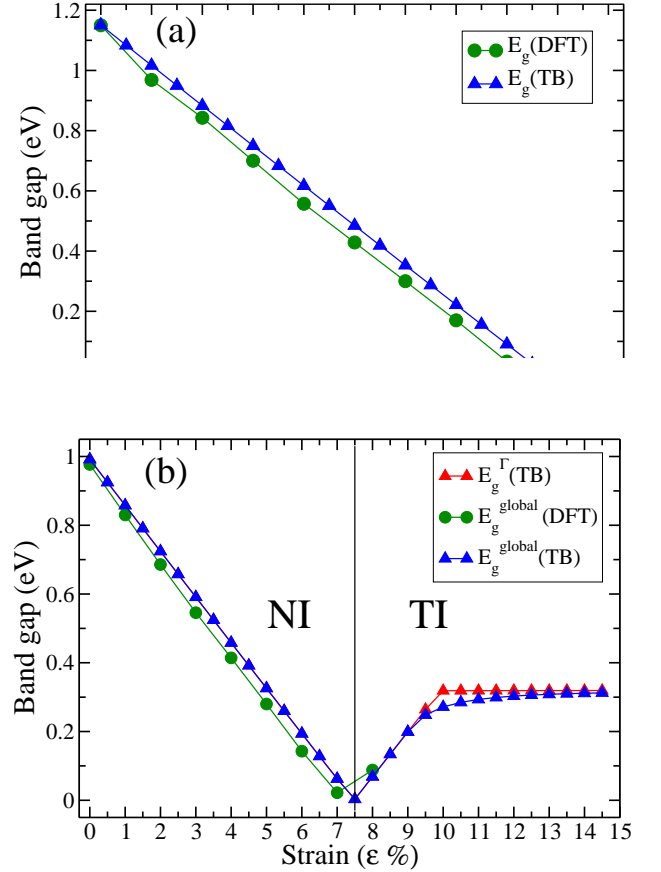
 
\centering
\vspace{20pt} 
\includegraphics[width=0.45\textwidth]{gap_strain.eps}
\includegraphics[width=0.46\textwidth]{gap_strain_SOC.eps}
\caption{The calculated band gaps of ML GeSnH$_2$ at the $\Gamma$ point ($E_g^\Gamma$) and the global energy gap ($E_g^{global}$) as a function of biaxial strain 
(a) without SOC, and (b) with SOC. The DFT data are taken from~\cite{ma2015}.}
\label{gap-strain}
\end{figure}

In Figs.~\ref{gap-strain}(a) and \ref{gap-strain}(b), we show the variation of the energy gap of ML GeSnH$_2$ as a function of biaxial
tensile strain without and with SOC, respectively. As shown in Fig.~\ref{gap-strain}(b), by applying biaxial tensile strain
in the presence of SOC, the band gap that is located at the $\Gamma$ point decreases steadily and eventually a band inversion occurs
at the critical value of 7.5\% strain.
By increasing the strain the band gap reaches $\sim$ 134 meV at a reasonable strain of 8.5\%. With increasing strain even further to 9.5\%, the band gap with SOC becomes indirect. As shown in TI phase in Fig.~\ref{gap-strain}(b), both $E_g^\Gamma$ and $E_g^{global}$ (global band gap) remain relatively unaffected by increasing the strain beyond 13\%.

Remarkably, the value of the band gap is significantly larger than $k_BT$
at room temperature ($\sim$ 25~meV), and large enough to realize the QSH effect in ML GeSnH$_2$ even at room temperature.
The excellent agreement between the results of our TB model and the DFT calculations in predicting the band inversion in this system upon biaxial 
tensile strain, implicitly confirms the validity of our proposed TB model.


\begin{figure}[ht]
\centering
\vspace{20pt}
\includegraphics[width=0.45\textwidth]{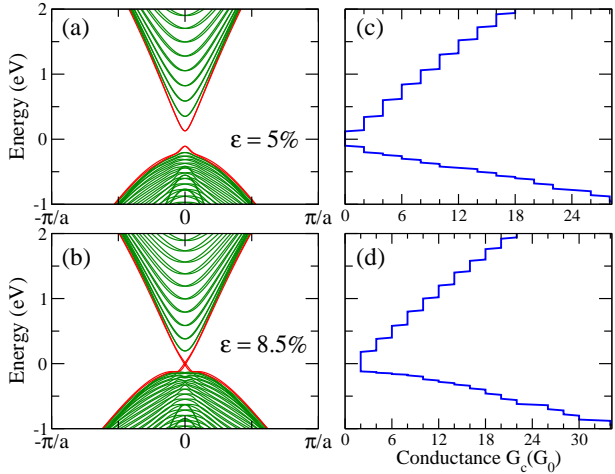}
\caption{The 1D energy bands and the total conductance (in units of $G_0=e^2/h$) of z-GeSnH$_2$-NR for $N=26$ in the presence of
(a) and (c) 5\%, (b) and (d) 8.5\% biaxial tensile strain.}
\label{band_zig}
\end{figure}

\section{\label{Phase}Topological phase transition of monolayer G\lowercase{e}S\lowercase{n}H$_2$ under strain} 
Using our TB model including spin orbit interaction, we showed that ML GeSnH$_2$ is a NI with a direct band gap of 0.977 eV. We showed that applying biaxial tensile strain modifies its electronic spectrum and a band inversion is taking place at $\epsilon=7.5\%$.

One way to show that there is a topological phase transition at the critical strain of $\epsilon=7.5\%$, is to perform calculations of the 1D band structure of nanoribbons with zigzag edges in the presence of biaxial tensile strain, and verify the existence of gapless helical edge states. Here, we study the edge state energy bands by cutting a 2D film of ML GeSnH$_2$ into nanoribbons. In our calculations, the Sn and Ge atoms at the edges of the nanoribbon are passivated by hydrogen atoms.
 
The width of a zigzag GeSnH$_2$ nanoribbon (z-GeSnH$_2$-NR) is defined by N, which is the number
of zigzag chains across the ribbon width. In order to reduce the interaction between
the two edges, the width of the nanoribbons is taken to be at least 10~nm.

We calculated the band structure of the corresponding nanoribbons.  
Note that the change of the hopping parameters and the on-site energies of the edge Ge or Sn atoms caused by the passivation
procedure is negligible. Therefore, we can use the results in Table~\ref{table1} and~\ref{table2} for the on-site energies of Ge and Sn atoms
and also for the hopping parameters corresponding to Ge-Sn bonds in unstrained or strained systems.
We have two different on-site energies of hydrogen atoms $E_{H_0}^s$ and $E_{H_1}^s$, which are pertinent
to the Hydrogen atoms that are introduced to passivate the Ge and Sn atoms on each edge, respectively.
We can write the hopping parameters related to the H-Ge and H-Sn bonds as
\begin{equation}
t_{H,X}^{ss}=V_{H,X}^{ss},~ t_{H,X}^{sp_{y}}=\pm V_{H,X}^{sp},~ (X=Ge,~Sn),
\end{equation}
where $+(-)$  denotes the lower (upper) H-X edge bonds. The numerical value of the mentioned hopping parameters and on-site energies of the hydrogen atoms can be obtained by a fitting procedure. The results are shown in Table~\ref{table3}. 

\begin{table}
\caption{Numerical values of the on-site energy of Hydrogen atoms and the hopping parameters related to the H-Ge and H-Sn bonds.
The energy is in units of eV.}
\label{table3}
\begin{ruledtabular}
\begin{tabular}{cccccc}
\textrm{$V^{ss}_{H,Ge}$}&
\textrm{$V^{sp}_{H,Ge}$}&
\textrm{$V^{ss}_{H,Sn}$}&
\textrm{$V^{sp}_{H,Sn}$}&
\textrm{$E^{s}_{H_0}$}&
\textrm{$E^{s}_{H_1}$}\\
\colrule
-4.87& 2.05 & -4.21 & 3.63 & -1.84 & -2.40
\end{tabular}
\end{ruledtabular}
\end{table}

Figs.~\ref{band_zig}(a) and~\ref{band_zig}(b) show the energy bands of z-GeSnH$_2$-NR with $N=26$ in the presence of 5\% and 8.5\%  
biaxial tensile strains, respectively. 
Gapless conducting edge bands are seen for strain of $\epsilon=8.5\%$. This is consistent with our proposal for the topological phase transition at $\epsilon=7.5\%$.  

It is now well established that for all time reversal invariant 2D band insulators a change in the $\mathbb{Z}_2$ topological invariant from zero to one, indicates a topological phase transition from a NI phase to TI. 
In our previous works\cite{sisakht2016strain,rezaei2017GeCH3}, we successfully used the algorithm of Fukui and Hatsugai~\cite{fukui1} in order to calculate
the $\mathbb{Z}_2$ number to characterize the topology of the energy bands. In this work, we implemented 
the same procedure to confirm the existence of two distinct topological phases in the electronic properties of ML GeSnH$_2$.
We found that the value of $\mathbb{Z}_2$ switches from zero to one at the critical strain of $7.5\%$, which confirms the topological nature of the phase transition in the electronic properties of ML GeSnH$_2$.

\section{\label{Transport}Electronic Transport in disordered G\lowercase{e}S\lowercase{n}H$_2$ Nanoribbons}
Transport measurement is a different approach to confirm the existence of helical 
gapless edge states, which is an important signature of the TI phase.
Therefore, we next study the transport properties of ML GeSnH$_2$ nanoribbons in the presence of strain
To this end, we calculate the conductance of z-GeSnH$_2$-NR using the Landauer formalism~\cite{datta1995,datta2005}. 
As is standard the z-GeSnH$_2$-NR is divided into three regions; the left and right leads and the middle scattering region. We initially assume the ribbon
to be perfect in order to satisfy the condition of ballistic transport.
In the Landauer approach, the total conductance $G_c(E)$ per spin of a nanoscopic device at the Fermi energy $(E_F)$ is given by
\begin{equation}
  G_{c}(E)=(\frac{e^2}{h})Tr[\Gamma _L(E)G_D^R(E)\Gamma _R(E)G_D^A(E)],
\end{equation} 
where $\Gamma _{L(R)}=i[\Sigma _{L(R)}(E)-\Sigma^\dagger_{L(R)}(E)]$, with $\Sigma_{L(R)}(E)$ being the self energy of the left (right) lead. 
$G_D^R(E)$ is the retarded Green's function of the device and $G_D^A(E)= {G_D^R}^\dagger(E)$.
The retarded Green's function of the device is given by
\begin{equation}
 G_D^R(E) = {[E-H_D-\Sigma_L^R(E)-\Sigma_R^R(E)]}^{-1}.
\end{equation}
Here $H_D$ is the Hamiltonian for the device region.
The numerically calculated total conductance, in units of $G_0={e^2}/{h}$, for $N=26$ as a function of energy is shown in Figs.~\ref{band_zig}(c) and \ref{band_zig}(d) in the
presence of 5\% and 8.5\% biaxial tensile strain, respectively.
For strains less than $7.5\%$ ($\epsilon<7.5\%$), the z-GeSnH$_2$-NR has nonzero conductivity only above a threshold energy corresponding to the
minimum energy of the conduction band, which opens a conducting channel with conductance in steps of $2G_0$. 
As shown in Fig~\ref{band_zig}(b), the total conductance at zero energy changes from 0 to 2 by applying biaxial tensile
strains larger than $7.5\%$ 
($\epsilon>7.5\%$). The non-zero conductance in the gap originates from the zero energy edge states and also indicates a topological phase transition from NI to TI
in ML GeSnH$_2$. These conducting edge states are protected by TRS leading to the
robustness of the electronic quantized conductance against backscattering by disorder and therefore holds great promise for spintronics
applications~\cite{gusev2011disorder, van2017effect}.

It would be helpful to further consider the effect of disorder on the electronic transport properties of this system. The disorder may be
originated from unwanted dislocations or other defects. Such calculations are another proof for our proposal of strain-induced
TI transition. In TIs, the edge states are robust against weak disorder, and only strong disorder can affect the electronic
properties.

\begin{figure}[ht]
\centering
\vspace{20pt}
\includegraphics[width=0.45\textwidth]{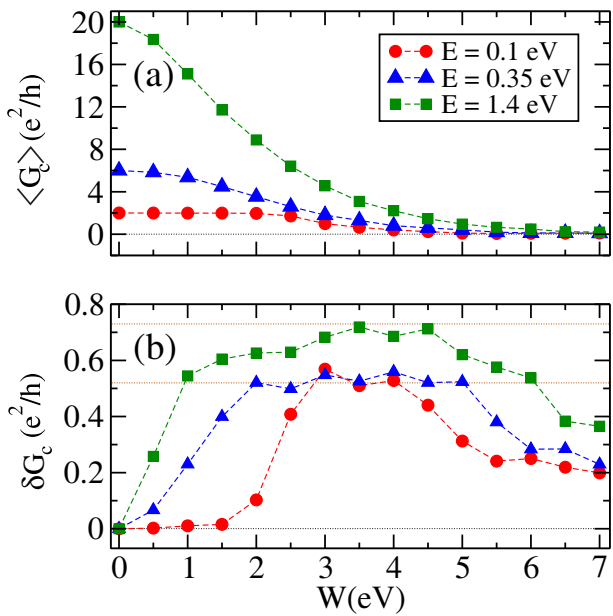}
\caption{(a) The mean conductance $\langle G_{c}\rangle$ and (b) the standard deviation of conductance $\delta G_{c}$ of z-GeSnH$_2$-NR with
$N=34$ (13~nm width) and 51~nm length as a function of disorder strength $W$.}
\label{Conduct_Devi}
\end{figure}

Here we introduce the disorder in our TB Hamiltonian model $H_{TB}=H_0+H_{SOC}$ using the so-called Anderson disordered
model~\cite{lee1981anderson} as
\begin{equation}
H_{D}=\sum_{i,\alpha}W_{i} c_{i,\alpha}^{\dag} c_{i,\alpha},
\label{HSOC}
\end{equation}
where $W_i$ is a random number uniformly distributed over the range $[-\frac{W}{2}, \frac{W}{2}]$. 
We assume a disordered z-GeSnH$_2$-NR with $N=34$ ($\sim$ 13 nm width) and 51~nm length as the middle region which is sandwiched between two
semi-infinite perfect leads
and calculate the conductance averaged over 100 different realizations. The width of the ribbon is chosen large enough in order to avoid finite-size effects.
Fig.~\ref{Conduct_Devi}(a) shows the average conductance of z-GeSnH$_2$-NR as a function 
of disorder strength $W$ in the presence of biaxial strain $\epsilon=8.5\%$ at three different values of the Fermi energy. The energy
$E_F=0.1$ eV corresponds to the energy in the middle of the band gap, while $E_F=0.35$ eV and $E_F=1.4$ eV are two energies in the bulk. 
It can be seen that the averaged quantized conductance at $E_F=0.1$ eV, which originates from the gapless edge states is insensitive to weak
disorder. The mean conductance decreases only for strong disorder of strength $W>1.5$ eV which is a signature of a topological insulator.

The universal conductance fluctuations (UCF)~\cite{lee1985UCF} are a mesoscopic phenomena, which is caused by the quantum interference of
electrons. The UCF are of order ${e^2}/{h}$ and correspond to the deviation of the conductance from its ballistic value $G=nG_0$.
The standard deviation of conductance in the diffusive regime and zero temperature depends only on the dimensionality and universality
class of the disordered mesoscopic system and is independent of the details of the system such as the disorder strength $W$, the conductance
$G_c$ and the sample size.
The UCF takes universal values for three types of symmetry classes corresponding to
$\beta$ = 1, 2 and 4, respectively.
In systems that preserve TRS and spin rotational symmetry (SRS), the symmetry index $\beta =1$ (orthogonal ensemble); and
$\beta =2$  corresponds to the case that TRS is broken (unitary ensemble); while $\beta =4$ for systems that preserve TRS but with broken SRS (symplectic ensemble)~\cite{hu2017numerical,hsu2018conductance}.
We study numerically the conductance fluctuations in the disordered z-GeSnH$_2$-NR in the presence of biaxial tensile strain.
The standard deviation of the conductance $\delta G_{c}={\langle {~(G_{c}-\langle G_c\rangle )}^2\rangle}^{\frac{1}{2}}$ of disordered
z-GeSnH$_2$-NR is plotted as a function of disorder strength $W$ at three different Fermi energies in Fig.~\ref{Conduct_Devi}(b).
There are no conductance fluctuations for energy $E_F=0.1$ eV in case of weak disorder strength, revealing the robustness of the helical edge states
against disorder in the QSH phase.
The $\delta G_c$ approach the value $\sim 0.52~e^2/h$, which corresponds to the UCF for the symmetry class $\beta=2$. Our total model Hamiltonian preserves TRS and the reason that the UCF shows the symmetry class $\beta=2$ is the following: since there are no spin-flip terms (Rashba SOC term) in our model Hamiltonian, we can block diagonal the Hamiltonian matrix with respect to the spin degrees of freedom with zero off-diagonal terms. Then we are dealing with two isolated and identical Hamiltonians corresponding to spin up and spin down states. Each of these blocks lack TRS due to the intrinsic SOC terms, and consequently the total UCF follows the $\beta=2$ symmetry class. A similar discussion can be found in Refs.~\cite{hsu2018conductance,choe2015universal}, except that in Ref.~\cite{hsu2018conductance} spin is not a good quantum number and the hamiltonian was block diagonalized with respect to the sublattice degrees of freedom.
For high energies and strong disorder strengths, the standard deviation approaches the value $\sim 0.73~e^2/h$, which belongs to the 
symmetry class $\beta =1$. We will compare the localization length with the spin relaxation length (spin relaxation length originates from the intrinsic SOC terms)~\cite{kaneko2010symmetry}. As long as the disorder is weak and the localization length is much larger than the spin relaxation length the SOC is significant and the system lacks SRS. But when disorder is strong enough and the localization length is much smaller than the spin relaxation length, we can ignore the SOC terms and SRS is preserved and the system follows the $\beta=1$ symmetry class. Finally, beyond $W\sim 6$ eV, where $\langle G_c\rangle$ approaches the value of $0.3~e^2/h$ the conductance fluctuations decreases and approaches the superuniversal  curve \cite{qiao2010universal} that is independent of dimensionality and symmetry. The superuniversal curve is beyond $W=7$ eV, which is not shown in Fig.~\ref{Conduct_Devi}.

\section{\label{Conclusion}Conclusion}
In summary, we constructed an effective TB model without and with SOC for ML GeSnH$_2$, which is able to reproduce
the electronic spectrum of this system in excellent agreement with the DFT results near the Fermi level.
Including SOC decreases the band gap from 1.155 eV to 0.977 eV.
In the presence of biaxial tensile strain, our proposed TB model predicts correctly the evolution of the band spectrum 
and also predicts a topological phase transition from NI to TI phase in ML GeSnH$_2$ at 7.5\% biaxial
tensile strain.
The global bulk gap, which is topologically protected, is 134 meV at a reasonable
strain of 8.5\%. This bulk gap exceeds the thermal energy at room temperature and is large enough to make ML GeSnH$_2$ suitable 
for room-temperature spintronics applications. 

The strong SOC and the applied mechanical strain are two essential factors that induce the topological phase transition from NI to QSH
phase in ML GeSnH$_2$. 
More interestingly, ML GeSnH$_2$ is a strain-induced TI with inversion asymmetry which makes it a
promising candidate for understanding intriguing topological phenomena like magneto-electric effects.
The TI nature of ML GeSnH$_2$ for strain $\epsilon>7.5\%$ was confirmed by calculating the $\mathbb{Z}_2$ 
topological invariant. Also we showed the existence of topologically protected gapless edge states in 
a typical z-GeSnH$_2$-NR in the presence of biaxial strain $\epsilon>7.5\%$.

In addition we found topologically protected gapless edge states in a typical z-GeSnH$_2$-NR for a biaxial
strain of $\epsilon=8.5\%$ in the presence of disorder by calculating electronic transport.
The conductance fluctuations reach the universal value of the unitary class $\beta=2$. For high Fermi energies and strong disorder the conductance fluctuations follow the orthogonal ensemble $\beta=1$. 

Acknowledgements: This work was supported by the FLAG-ERA project TRANS-2D-TMD.
 
 \nocite{*}

\bibliography{GeSnH2}

\end{document}